\documentstyle[12pt]{article}
\newcommand{\de}{\delta}

\newcommand{\eq}[1]{eq.\hspace*{.1em}(\ref{#1})}
\newcommand{\eqs}[1]{eqs.\hspace*{.1em}(\ref{#1})}

\newcommand{\GeV}{\,\mbox{GeV}}
\newcommand{\MeV}{\,\mbox{MeV}}

\def\lsim{\mathrel{\rlap{\lower3pt\hbox{\hskip0pt$\sim$}}
    \raise1pt\hbox{$<$}}}         %less than or approx. symbol
\def\gsim{\mathrel{\rlap{\lower4pt\hbox{\hskip1pt$\sim$}}
    \raise1pt\hbox{$>$}}}         %greater than or approx. symbol

\newcommand{\matel}[3]{\langle #1|\,#2\,|#3\rangle}

\def \as{\relax\ifmmode\alpha_s\else{$\alpha_s${ }}\fi}
\def\MSbar{\relax\ifmmode\overline{\rm MS}\else{$\overline{\rm MS}${ }}\fi}

\def\baeq{\begin{appeq}}     \def\eaeq{\end{appeq}}
\def\baeeq{\begin{appeeq}}   \def\eaeeq{\end{appeeq}}
\newenvironment{appeq}{\beq}{\eeq}
\newenvironment{appeeq}{\beeq}{\eeeq}

\newcounter{hran}
\renewcommand{\thehran}{\thesection.\arabic{hran}}

\def\bmini{\setcounter{hran}{\value{equation}}
\refstepcounter{hran}\setcounter{equation}{0}
\renewcommand{\theequation}{\thehran\alph{equation}}\begin{eqnarray}}
\def\bminiG#1{\setcounter{hran}{\value{equation}}
\refstepcounter{hran}\setcounter{equation}{-1}
\renewcommand{\theequation}{\thehran\alph{equation}}
\refstepcounter{equation}\label{#1}\begin{eqnarray}}

\def\emini{\end{eqnarray}\relax\setcounter{equation}{\value{hran}}\renewcommand{\theequation}{\thesection.\arabic{equation}}}

\def\ga{\mathrel{\mathpalette\fun >}}
\def\fun#1#2{\lower3.6pt\vbox{\baselineskip0pt\lineskip.9pt
  \ialign{$\mathsurround=0pt#1\hfil##\hfil$\crcr#2\crcr\sim\crcr}}}

\def\half{{\textstyle {1\over2}}}

\def \al {\relax\ifmmode{\alpha}\else{$\alpha${ }}\fi}
\def \be {\relax\ifmmode{\beta}\else{$\beta${ }}\fi}
\def\ga{\gamma}

\def\ben{\begin{enumerate}}  \def\een{\end{enumerate}}
\def\bit{\begin{itemize}}    \def\eit{\end{itemize}}
\def\beq{\begin{equation}}   \def\eeq{\end{equation}}
\def\beeq{\begin{eqnarray}}  \def\eeeq{\end{eqnarray}}
\newskip\humongous \humongous=0pt plus 1000pt minus 1000pt
\def\caja{\mathsurround=0pt}

\newif\ifdtup

\def\eqal2#1{\,\vcenter{\openup1\jot
\caja   \ialign{\strut \hfil$\displaystyle{##}$&\hfil$
\displaystyle{{}##}$\hfil &$
\displaystyle{{}##}$\hfil\crcr#1\crcr}}\,}

\begin{document}

\begin{titlepage}
\renewcommand{\thefootnote}{\fnsymbol{footnote}}

\begin{flushright}
CERN-TH/96-40\\
hep-ph/9602324
\end{flushright}
\vspace{.3cm}
\begin{center} \LARGE
{\bf On the Problem  of Boosting \\
Nonleptonic $b$ Baryon Decays}
\end{center}
\vspace*{.3cm}
\begin{center} {\Large
N.G. Uraltsev\\
\vspace{.4cm}
{\normalsize 
{\it TH Division, CERN, CH-1211 Geneva 23,
Switzerland}\\
and\\
{\it St.Petersburg Nuclear Physics Institute,
Gatchina, St.Petersburg 188350, Russia}\footnote{Permanent address}\\
\vspace{.3cm}
\vspace*{.4cm}
}}

{\vspace*{.2cm}\Large{\bf Abstract}\\}
\end{center}
\begin{quote}
The constituent picture of hadrons implies certain quantum
mechanical inequalities which must hold in the potential models. Basing on 
this qualitative consideration I argue that it is not easy to 
increase significantly the scale of the flavour-dependent $1/m_b^3$
effects within the heavy quark expansion preserving the conventional
constituent picture of heavy flavour hadrons. I briefly address the physical
consequences one might expect if the effects of weak scattering and
interference are attempted to be pushed above the $10\%$ level within $1/m_b$
expansion not invoking qualitatively different mechanisms including 
violations of duality.

\end{quote}
\vspace*{\fill}
CERN-TH/96-40\\
February 1996
\end{titlepage}
\addtocounter{footnote}{-1}

\newpage

Heavy quark expansion allows one to address systematically the inclusive widths
of the heavy flavour hadrons based genuinely on QCD with minimal -- though
rather important -- qualitative information supplied by experiment about the
behaviour of QCD in the strong interaction regime. The essential elements of the
present theoretical technology were set up in mid 80's \cite{vs} and were
applied already then to estimate the preasymptotic effects in charmed and 
beauty particles. Later, the systematic study of the $1/m_Q$ expansion for the
inclusive decay rates has been done with special attention to the
subtleties involved in the application of the OPE to the Minkowsky decay
processes. In particular, it was shown \cite{gluon,we} that the power 
corrections
to the inclusive widths (both semileptonic and nonleptonic, as well as
radiative ones) are absent at the $1/m_Q$ level and start with $1/m_Q^2$
terms. These leading corrections do not depend explicitly on the flavour of the
spectator; they were calculated in \cite{we} (for the review see
\cite{stone}). Although these leading effects differentiate lifetimes of mesons
and baryons, their effects are not large in the individual parton level 
decay channels of $b$ particles, and appear to be additionally suppressed
numerically in the total decay width.

The flavour-dependent effects emerge at the $1/m_b^3$ level but are numerically
enhanced and, in general, constitute several per cent of $\Gamma_{\rm tot}$ 
in beauty. They are given
by the expectation values of the four-fermion operators
\beq
O_{\alpha\beta}\;=\;\bar b \ga_\al(1\!-\!\ga_5) q \:\bar q \ga_\be 
(1\!-\!\ga_5)
b \qquad \;;
\;\;\qquad \qquad \qquad  O\;=\;O_{\al\al}\qquad \qquad \qquad \qquad 
\label{1}
\eeq            
(with two possible colour contraction schemes), where $q$ is the appropriate
light quark. The Lorentz scalar operators $O$ emerge when one integrates out
the diquark loop and describe interference (PI) in the
decays of B mesons as well as weak scattering (WS) in baryons;  the different
combination of the components of $O_{\al\be}$ results due to $q\bar q$
intermediate pair and is responsible for weak 
annihilation (WA) in mesons and PI in baryons \cite{vs}. The tree level
coefficient functions are well known in the general case:
$$
c^{qq}\;=\;\frac{G_F^2 P^2}{2\pi}\,|\mbox{KM}|^2 \; 
p\left(1-\frac{m_1^2+m_2^2}{P^2}\right)\;\;,
\qquad \;\;
c^{q\bar q}_{\al\be}\;=\;-\frac{G_F^2 P^2}{6\pi}\,|\mbox{KM}|^2\; 
\left(A\de_{\al\be}-B\frac{P_\al P_\be}{P^2}   \right)
$$
$$
A\;=\;p\left(1-\frac{m_1^2+m_2^2}{2P^2} -\frac{(m_1^2-m_2^2)^2}{2P^4}\right)
\;\;,
\qquad \;\;\; 
B\;=\;p\left(1+\frac{m_1^2+m_2^2}{P^2} -\frac{2(m_1^2-m_2^2)^2}{P^4}\right)
$$
\beq
p\;=\;\left[ \left(1-\frac{(m_1+m_2)^2}{P^2}\right) \left( 
1- \frac{(m_1-m_2)^2}{P^2}\right)  \right] ^{\half}
\label{2}
\eeq
where $m_1$, $m_2$ are the quark masses in the loop, $P$ is the total momentum
flowing into it (normally identified with the momentum of the heavy quark
$p^b$ or of the hadron $P^H$) and $|\mbox{KM}|^2$ symbolically denotes the
product of the quark mixing angles. The QCD corrections to the coefficient 
functions
are also known \cite{vs} and include colour traces depending on 
$N_c$ and $c_\pm$ and, in particular, the so-called
``hybrid'' renormalization coming from the scales below $m_b$. The corrections
to the width are given \cite{vs,gluon,WA} by the forward matrix element of 
the corresponding
generic sum $c_i\cdot O_i$ over the particular hadron, $B$, 
$\Lambda_b$ etc.:
\beq
\Delta \Gamma_{H_Q}\;=\;\frac{1}{2M_{H_Q}}\,
\matel{H_{Q}} {c_i\cdot O_i}{H_{Q}}\;\;.
\label{2a}
\eeq

The situation with the matrix elements is less clear and is the subject of the
present discussion. In the case of mesons one employs factorization:
\beq
\matel{B_{q'}} {\bar b \ga_\al(1\!-\!\ga_5) A^a q \:\bar q \ga_\be 
(1\!-\!\ga_5) A^b b}
{B_{q'}}\;=\; f_B^2 P^B_\al P^B_\be \cdot \frac{1}{9}{\rm Tr}\,A^a\,
{\rm Tr}\,A^b\,\de_{q'q}
\label{3}
\eeq
where $A^{a,b}$ are colour matrices and $q$, $q'$ are light quark flavours. WA
appears to be strongly suppressed {\em in the factorization approximation} by 
the ratio $m_c^2/m_b^2$ (for the dedicated discussion see \cite{gluon,WA}).

The baryonic matrix elements are even less certain. Their estimates rely so far
mostly on simple potential quark models of heavy flavour baryons. Ignoring for
a moment the colour indices, one uses the Fierz identities and the equation of
motion for the $b$ field to write \cite{vs}
\beq
O\;=\;\bar b \ga_\al(1\!-\!\ga_5) b \;\bar q \ga_\al (1\!-\!\ga_5) q\;\;;\;
\qquad P_\al P_\be \;O_{\alpha\beta}\;=\;-\frac{1}{2}\,P^2\;
\bar b \ga_\mu(1\!+\!\ga_5) b \;\bar q \ga_\mu (1\!-\!\ga_5) q \;\;.\; 
\label{4}
\eeq
In the current $\bar b \ga_\al(1\!-\!\ga_5)b$ the vector part is nonzero for
$\al=0$ and the axial part survives for the spacelike components. If the colour
singlet $b$ quark current is considered, the former represents the heavy quark
density, whereas the latter is the $b$ quark spin density which decouples from
the light degrees of freedom as
$m_b$ goes to infinity. Therefore, in the matrix element over the $\Lambda_b$
state the axial current does not contribute in the heavy quark limit \cite{vs}
(it does for $\Sigma_b$ and $B$ states). Strictly speaking, this general
statement holds only for the particular colour structure of the four-fermion
operator.

Further simplification arises when one applies the 
description of the baryon relying on ordinary quantum mechanics of only the 
constituent quarks. Then the single (antisymmetric) 
colour structure survives and one has to know the
unique matrix element 
\beq
\frac{1}{2M_{H_Q}}
\matel{\Lambda_b} 
{\bar b^i \ga_0 b^i \:\bar q^j \ga_0 q^j} {\Lambda_b}\;
=\;|\Psi^d(0)|^2
\label{5}
\eeq
where $\Psi^d$ denotes the heavy-light diquark wave function. Collecting all
coefficients together (see, e.g., \cite{vs,Bigi}) one arrives at the following
expressions for the effects in $\Lambda_b$:
\beq
\frac{\Gamma_{WS}}{\Gamma_0}\;\simeq \; 96\pi^2 c_-^2\:
\frac{|\Psi^d(0)|^2}{m_b^3}
\label{6}
\eeq
\beq
\frac{\Gamma_{PI}}{\Gamma_0}\; \simeq \; -96\pi^2 c_+(c_- -\frac{c_+}{2}) \:
\frac{|\Psi^d(0)|^2}{m_b^3}\;\;;
\label{7}
\eeq
$\Gamma_0$ denotes the (phase space uncorrected) bare semileptonic width
$\Gamma_0=G_F^2 m_b^5|V_{cb}|^2/(192\pi^3)$ and $c_\pm$ are the standard short distance
coefficients. I neglected here minor corrections due to the final state quark 
masses, \eq{2}, which anyway do 
not exceed the effect of the higher dimension operators,
and more essential hybrid renormalization effects (to be briefly addressed 
later). These expressions are the standard starting point~\cite{vs} for the 
numerical evaluation of the flavour-dependent preasymptotic corrections.

Before proceeding to the specific subject of the current paper, let me note
that the factorization \eq{3} clearly holds in the constituent quark ansatz,
with \cite{AzShur}
\beq
f_B^2\; =\;12 \,\frac{|\Psi(0)|^2}{M_B}
\label{8}
\eeq
where $\Psi(0)$ now denotes the light quark wavefunction at zero separation and 
the factor $12$ comes from the colour and spin traces.

The wavefunctions are governed  by the strong
interaction dynamics and thus seem to be very uncertain, depending crucially on
the quark interaction even within the potential description. Therefore, it is
tempting to allow for the larger effects of PI and WS in $\Lambda_b$
pushing $|\Psi^d(0)|^2$ up to meet the (not firmly established yet, though)
experimental evidence \cite{exp} 
that the $\Lambda_b$ lifetime can be noticeably smaller
than that of $B$. I will argue that such an option would require certain
revision of the simple constituent models for heavy baryons, in particular,
applicability of the heavy flavour symmetry at a quantitative level.

In the QM description one has the following constraints on the wavefunctions:
\beq
\Psi(0)\;=\; \int\, \frac{d^3\,p}{(2\pi)^3} \:\Psi(p)
\label{9}
\eeq
\beq
\int\, d^3\,x\:|\Psi(x)|^2\;=\;\int \,\frac{d^3\,p}{(2\pi)^3}\:|\Psi(p)|^2\;=\;
N\;=\;1\;\;.
\label{10}
\eeq
One must also assume that $p^2|\Psi(p)|^2$ falls off rapidly above certain 
characteristic hadronic scale $\mu$, for the physics of the harder modes is
absorbed into the coefficient functions of the effective low energy operators.
Assuming for simplicity that $\Psi(p)=0$ at $|\vec{p}\,| >\mu$, one uses the
Cauchy-Bunyakowski-Schwartz inequality to get
\beq
|\Psi(0)|^2\;\le \; \int_{\vec{p}\,^2<\mu^2}\, \frac{d^3\,p}{(2\pi)^3} 
\;\cdot N \;=\; \frac{\mu^3}{6\pi^2}\;\;,\;\;\;\;\;
f_B^2\;\le \;\frac{2\mu^3}{\pi^2M_B}\;\;.
\label{11}
\eeq
Thus there is a simple upper bound on the wavefunction in terms of the phase
space allocated for the system. The inequality saturates when $\Psi(p)=\rm
const$ which yields the  ``finite size $\delta$-function''
$\Psi(x)=\mu^{3/2}\frac{1}{\pi}\sqrt{\frac{3}{2}}\,
(\sin({\mu |x|})/(\mu |x|)-\cos{\mu |x|})/(\mu^2 x^2)$ 
in the coordinate space.

Inequality (\ref{11}) appears to be rather restrictive: taking $\mu$ as large as
$1\GeV$ one has 
\beq
f_B \lsim 200\MeV \;\;,\;\;\;f_D \lsim 330\MeV\;\;.
\label{13}
\eeq
This upper bound for $f_B$ is close to the existing theoretical estimates and
maybe even somewhat lower than the expected values for $f_{B_s}$. It is worth
noting the important role of the colour factor $N_c=3$ in $f_B^2$ in
accommodating such, naively quite moderate, magnitude of $f_B$. Eq.~(\ref{11})
demonstrates that, for the dimensional estimates, the quantity $\pi f_B$ rather
than plain $f_B$ gives the proper scale; this strongly offsets, for example,
apparently huge enhancement factor $24\pi^2$ of the two body phase space in PI,
WA or WS compared to the three-body one in the free decay kinematics.

Adopting $\mu=1\GeV$ as the reference point we obtain numerically
\beq
|\Psi^d(0)|^2\;\lsim \; 0.017\GeV^3\;\;.
\label{14}
\eeq
This estimate thus sets the scale of what seems to be the maximal ``natural''
diquark density. It is justified to state that the values significantly larger
than this and, correspondingly, the related $1/m_b^3$ corrections to the
inclusive widths, call for a specific underlying mechanism to be added to the
conventional picture.

One such dynamical mechanism is known -- the perturbative short distance 
enhancement in $B$ mesons due to the hybrid gluon exchange \cite{vs}:
\beq
f_B^2=\left[ \frac{\as(\mu^2)}{\as(m_b^2)} \right]^{4/b} \cdot 
\,12\,\frac{|\Psi(0)|^2}{M_B}\;\;,\;\;\;\;\;\; 
b=\frac{11}{3}N_c-\frac{2}{3} n_f\simeq
9\;\;.
\label{20}
\eeq
Although for large $\mu$ this perturbative factor cannot be large, it
literally enhances $f_B^2$ by only a factor of $1.3\div 1.5$ (I use here 
the $V$ scheme $\as$ as a physically adequate one), it is sufficient to
move $|\Psi(0)|^2$ in $B$ to the ``comfortable'' zone near $0.01\GeV^3$ and to
allow for a less extreme value of $\mu\gsim 0.8\GeV$, in particular adopting 
the theoretically
preferable values of $f_B\simeq 160 \MeV$ \cite{braun}.

\begin{figure}
\vspace{5.0cm}
\includegraphics{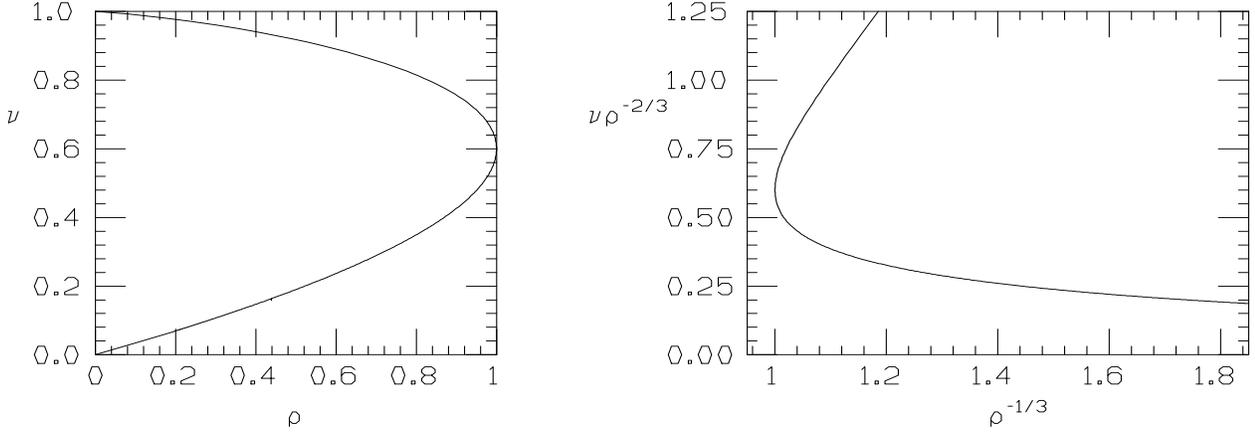}
\caption{{\bf Left:} 
possible values of $\;\;6\pi^2|\Psi(0)|^2\;\;$ ($\,\rho\,$) $\;\;$ and 
$\;\;\mu_\pi^2\;\;\;\;$ ($\,\nu\,$) $\;\;$ in units of $\mu$. \newline
{\bf Right:} 
the range of variation of the dimensionless ratio
$\mu_\pi^2/(6\pi^2|\Psi(0)|^2)^{2/3}$ versus the degree of saturation of
inequality (12).\newline 
The allowed regions lie between the upper and lower branches.
}
\end{figure}

Large scale of the essential momenta of the constituents in $B$ imply
relatively high $\mu_\pi^2$, the expectation value of the kinetic operator in
$B$ mesons: assuming, as before, that $\Psi$ vanishes above $\mu$ one gets, for
example, $\mu_\pi^2=3/5 \mu^2$ if $\mu$ is the minimal cutoff capable to
accommodate given $|\Psi(0)|^2$ and, therefore,
\beq
\mu_\pi^2\;=\;\frac{3}{5}\left(6\pi^2|\Psi(0)|^2\right)^{2/3} \;\simeq 
\;\frac{3}{5}\left(\frac{\pi^2}{2}\right)^{2/3} f_B^{4/3} M_B^{2/3} 
\left[ \frac{\as(m_b^2)}{\as(\mu^2)} \right]^{8/27} \;
\simeq \; 0.4\GeV^2\;\;;
\label{21}
\eeq
for larger $\mu$ somewhat smaller $\mu_\pi^2$ is also possible, although the
decrease cannot be dramatic unless $\mu$ is really large. In Fig.~1 I show the
allowed values of the dimensionless ratios
$$
\nu\;=\;\frac{\mu_\pi^2}{\mu^2}\;\;\;,\;\;\;\; 
\rho\;=\; \frac{6\pi^2|\Psi(0)|^2}{\mu^3}
$$
relevant for the absolute values of the hadronic parameters $\mu_\pi^2$ and
$f_B^2$. The range of the direct ratio of $\mu_\pi^2$ to 
$(6\pi^2|\Psi(0)|^2)^{2/3}$ which is given by 
$\nu\rho^{-2/3}$ is also interesting and shown 
in the right plot. 
Smaller values of $\mu_\pi^2$ are possible only beyond the two
particle picture of $B$ mesons.
Relaxing any constraint on $\mu$ one can have, in principle, arbitrary value of
$\nu \rho^{-2/3}$, i.e. in this case no lower bound on $\mu_\pi^2$
emerges. Still, one can obtain the $\mu$-independent lower bound on the 
expectation value of the fourth power of momentum, $\langle \vec p\,^4
\rangle\,$:
\beq
2^{1/2}3^{1/4}\,\pi \:\left(\langle \vec p\,^4\rangle\right)^{3/4}\;\ge\;
6\pi^2 |\Psi(0)|^2\;\;.
\label{21a}
\eeq
This follows from one of the Sobolev's family of inequalities occurring in the
so-called {\em embedding theorems}, namely, 
\beq
|f(y)|^2 \;\le \; \frac{1}{2^{1/2}3^{3/4}\pi} \;
\left(\int\,d^3\,x\:|f(x)|^2\;\right)^{1/4} \cdot 
\left(\int\,d^3\,x\:|\nabla^2f(x)|^2\;\right)^{3/4}\;\;.
\label{22a}
\eeq

We see that the QCD sum rule
determination of $f_B$ and $\mu_\pi^2$ \cite{braun} looks consistent 
from the above 
perspective; it is also in agreement with another, more rigorous QCD 
lower bound on the kinetic operator \cite{boundp} and with the 
more phenomenological estimates \cite{third}.

An attempt to boost $|\Psi^d(0)|^2$ in $\Lambda_b$ requiring larger $\mu$ may
seem not to imply necessarily the large expectation value of the kinetic
operator: the momentum of one of the light quarks can be balanced by another
light quark rather than by $b$ if the two light quarks are strongly correlated.
In other words, in $b$ baryons the moments of $|\Psi^d(p)|^2$ do not coincide
with the expectation values of the operators $\bar b (i \vec D)^k b$ appearing
in the $1/m_b$ expansion. In particular, 
the possibility may remain that $\mu_\pi^2(\Lambda_b)$ is small. 
Although this
would uniformly enhance the $\Lambda_b$ decay rate compared to $B$ due to the
Lorentz dilation \cite{dil}, this effect can hardly reach even the two
percent level.
On the other hand, the very same inequality (\ref{21a}) still 
holds for the diquark 
density~\footnote{
It can be proved in the same way as (\ref{22a}), or making use of inequality
(\ref{22a}) and the H\"{o}lder inequality 
$\int f\,g \le (\int f\,)^{1/4}
(\int f\,g^{4/3})^{3/4}$ valid for positive $f$ and $g$. Inequality
(\ref{21c}) is saturated when $\Psi^{\Lambda_b}(x;y)=f_0(x)\cdot \phi(x-y)$ 
where $f_0$ saturates (\ref{22a}) and $\phi$ is arbitrary.
}:
\beq
2^{1/2}3^{1/4}\,\pi \:\left(\langle \vec p_b^{\:4}
\rangle_{\Lambda_b}\right)^{3/4}
\;\ge\;
6\pi^2 |\Psi^d(0)|^2
\label{21c}
\eeq
with $|\Psi^d(0)|^2= \int d^3\,y \,|\Psi^{\Lambda_b}(0;y)|^2\;$ and 
$\;\langle \vec
p_b^{\:4}\rangle_{\Lambda_b}=\int d^3x\, d^3y\, 
|(\nabla_x\!+\!\nabla_y)^2\Psi^{\Lambda_b}(x;y)|^2$ even without $u
\leftrightarrow d$ symmetry
constraints on $\Psi^{\Lambda_b}$.

Let us briefly examine the consequences of the hypothesis that the essential
momenta of the light quarks are large and $\mu$ noticeably exceeds $1\GeV$.
I do not try to speculate here whether any QM-type potential model can be
formulated in a self-consistent fashion if the light quark momenta are of such
high scale; it is more reliable to discuss what would be the expected
model-independent features for heavy flavours from the general QCD perspective.
Clearly, it would destroy the applicability of the heavy quark expansion to the
corresponding charm hadrons, including the spectrum and exclusive formfactors. 
It is not natural to expect that in
such a case all traces of the symmetry relations are wiped away; rather,
one would think that 
the symmetry pattern still persists at a qualitative level,
whereas quantitative model-independent QCD predictions cannot be done. Some
static characteristics may, possibly, survive 
if, say, the two light quarks form a very compact colour-antitriplet
configuration which is only softly bound to the heavy quark and, therefore, 
there are two different hadronic scales in the problem. It is not
clear how natural this option is, but such a peculiarity must manifest
itself in a number of other processes. 

Large intrinsic momenta of the spectators would hardly justify the applicability
of the standard expressions for the spectator-dependent $1/m_Q^3$ inclusive 
corrections to the widths. It is most transparent in the case of interference:
the final expressions for the interference term via $|\Psi(0)|^2$ clearly
imply that the typical momenta of the light decay quarks are larger than the
momenta of the spectators -- and the former are typically about $1.5\GeV$ even
in beauty decays. Moreover, it is this ratio of the intrinsic to the final
state quark momenta that controls the importance of the higher order power
corrections and the significance of the ``exponential'' terms signaling the
duality violation. The trend of such effects can hardly be predicted {\em a
priori} in actual QCD.

While the intrinsic momenta of the spectators somewhat above $1\GeV$ can still
allow for semi-quantitative analysis of the nonleptonic decays in beauty, the
charm decay rates would at best offer a possibility to discuss only the
qualitative pattern, with all spectator-dependent corrections being not
suppressed at all. Even the semileptonic width of charmed particles can be
seriously affected if such a scenario represents reality.

Turning back to the effects of PI and WS in $\Lambda_b$, let us assume that
$\mu\simeq 1 \GeV$ and set, according to \eq{14}, the diquark density
$0.017\GeV^3$. We then get numerically
\beq
\;\;\;
\frac{\Gamma_{WS}}{\Gamma_{B_0}} \;\simeq \;0.067\;\;\;,\qquad \qquad \qquad
\qquad  
\frac{\Gamma_{PI}}{\Gamma_{B_0}} \;\simeq \;\;-0.028\;\;\;.\qquad \qquad \qquad
\label{22}
\eeq
The expressions for the net effect of the hybrid renormalization has been
given in the second paper \cite{vs}; using the adopted approximation for the
four fermion matrix elements the perturbative corrections read as follows:
$$
c_{WS}^{\Lambda_b} \;\simeq \;\frac{1}{2}\left[c_+^2 + c_-^2 +
\frac{1}{3}(1-\kappa^{9/2}) (c_+^2 -
c_-^2) \right]-\frac{c_+^2 - c_-^2}{2}\kappa^{9/2} -
\frac{4}{9}(c_+^2 - c_-^2) \kappa^{9/2}(\kappa^{-2}-1)
$$
$$
c_{PI}^{\Lambda_b} \;\simeq \;-\frac{1}{8}\,
\left\{  (c_+ + c_-)^2 + \left[ \frac{1}{3}
-\frac{4}{3} \kappa^{9/2}
\left(1+\frac{4}{3} (\kappa^{-2}-1) \right) 
\right]  (5c_+^2 + c_-^2- 6c_+c_-)\right\}
$$
\beq \kappa\;=\; \left[\frac{\alpha_s(\mu^2)}{\alpha_s(m_b^2)}\right]^{1/9}
\;\;\;.\qquad \qquad \qquad
\label{23}
\eeq
Applied literally, they amount to the 
additional factors $1.14$ for WS and
$1.05$ for PI (I used here the value of the strong coupling in the $V$ scheme
$\alpha_s(2.3\GeV)=0.336$ \cite{vol}). 
Needless to say, the accuracy of the estimates (\ref{22}) relying on the simple
model for the four-fermion matrix elements is not
high, nor even of their ratio $\Gamma_{\rm WS}/\Gamma_{\rm PI}$ 
from which the unknown
wavefunction naively drops out. Therefore it would be unjustified, in my
opinion, to state the significant cancellation which is literally
suggested by \eqs{22}.
It is more reasonable to expect merely that PI somewhat 
decreases the possible effect of WS in $\Lambda_b$ \cite{Bigi}, at least in the
framework of the leading terms in the $1/m_b$ expansion. Since there is no
large enough room for the (logarithmic) perturbative physics for large $\mu$, 
the effect of the
hybrid renormalization can be viewed, conservatively, as the uncertainty 
of the simple estimates which typically neglect such effects originating in 
the domain below $m_b$. 

Very recently the heavy-light diquark density at zero separation in $\Lambda_b$
was estimated
in the potential model-motivated way using the information on the hyperfine 
splitting in beauty mesons and $\Sigma$ states \cite{rosner}. The value
suggested by that consideration agreed with the numerical bounds I discussed 
above and, thus, the resulted effects of WS and PI were only at a few percent 
level. \vspace*{.15cm}

To summarize, I argued that the conventional constituent models of heavy
flavour hadrons have restrictive intrinsic limitations on the possible size of
the matrix elements governing 
$1/m_Q^3$ corrections to widths. These qualitative arguments
are quantified by inequalities (\ref{11}) and (\ref{21a}). They seem to be of
rather general nature and rely only on the assumption that the bound state
dynamics is governed by the soft field components; even the number of the
effective degrees of freedom does not seem to be crucial. The actual 
enhancement may thus come from the effects of shorter distances in various 
forms.
Attempts to boost the preasymptotic 
effects in the width in the straightforward manner above the scale of $10\%$  
in $\Lambda_b$ require an essential revision of the assumptions set up in such
models and would lead to important consequences, in particular, for the
applicability of the heavy flavour symmetry to charmed particles. One can also
guess that such localized hadrons' wavefunctions maybe not easy to extract 
reliably from the existing lattice simulations, requiring rather big lattice to
correctly reproduce both relatively large momenta and ``usual'' soft
components.

\vspace*{0.5cm}

\noindent
{\bf Acknowledgments:} \hspace*{.15em}
I am thankful to P.~Ball, I.~Bigi, V.~Braun, J.~Rosner and M.~Shifman for 
useful 
discussions and their interest in the theoretical problems of the lifetimes of 
$b$ particles, and to my mother Nina N. Uraltseva for discussing mathematical
aspects.

\vfill
\end{document}